# Physical Basis for Band Transport and Dimensionality in Amorphous Oxide Semiconductor Field-Effect Transistors

Ananth Dodabalapur,[1,2,*] Chankeun Yoon,[1,2] and Xiao Wang [1,2]

[1] *Chandra Family Department of Electrical and Computer Engineering, The University of Texas at Austin, Austin, Texas 78712, USA*

[2] *Microelectronics Research Center, The University of Texas at Austin, Austin, Texas 78758, USA*

\* *Author to whom correspondence should be addressed: ananth.dodabalapur@engr.utexas.edu*

## Abstract

**A consistent and widely accepted physical basis for interpretation of charge transport in amorphous oxide semiconductor (AOS) field-effect transistors (FETs), and more generally device physics, has been hampered by uncertainties in crystalline order, dimensionality, and the effects of a significant density of traps. The overarching theme of this paper is to build and justify a much-needed conceptual framework for describing advanced AOS transistors, particularly those with very small channel lengths. Combining new work and selecting prior research results on charge transport and device physics together with literature reports from various groups on morphology, physical properties, electronic structure and percolation effects, the main evidence that is available in support of a trap-influenced band transport picture in quasi-2-dimensional channels in high mobility AOS FETs is presented.**

## I.    Introduction

Field-effect transistors based on amorphous oxide semiconductors (AOS) are being actively investigated for continued use in display applications [1-3], as well as potential use in back-end-of-the-line circuitry in silicon integrated circuits. [4-7] Studies on charge transport physics have informed the construction of device models for simulation purposes.[8-16] Earlier reports on charge transport phenomena in AOS FETs mostly invoked hopping transport models, particularly variable-range hopping (VRH). [17,18] The higher mobilities in these semiconductors were accounted for by a larger delocalization length compared to lower mobility semiconductors in which VRH has been used to model charge transport. Multiple trapping and release (MTR) models have also been proposed by several groups to explain transport. [19-24] The MTR model has been quite successful at accounting for a range of phenomena observed in experimental results. [25-27]

Notwithstanding all these reports, the underlying physical justification in using band transport models and assumption of quasi-2-dimensional carrier confinement in the channel has not been explicitly made to an adequate degree. For both fundamental studies, and in device analysis, a comprehensive and unified discussion is necessary as to why these assumptions are appropriate for nominally amorphous semiconductors. The main goal of this paper is to bring together evidence from our own work on the derivation of density-of-states, mean free path estimates, scattering mechanisms, quantitative evaluation of various carrier velocity behavior at high electric fields, Hall effect measurements, and observations of metal-insulator transition with literature values of effective masses and other physical parameters, band structure parameters, and structural ordering on nanometer length scales. This combined suite of evidence indicates that charge transport in high-mobility AOS FETs is best understood, not with hopping transport models but with trap-influenced transport through extended states in a quasi-2-dimensional channel.

It is emphasized that there is relatively less data available on parameters such as effective mass, low and high frequency dielectric constants, and dominant phonon energies for charge transport in AOS, especially compositional dependent values. In many cases, estimates must be made based on reported values in crystalline binary semiconductors such as $Ga_2O_3$, $In_2O_3$, and ZnO. [28-30] This will naturally lead to uncertainties in calculations and the impact of this uncertainty is discussed. In the final section before the conclusions, a combined discussion on band transport and metal insulator transition in AOS FETs is presented. Mobility edge energies relative to the band edge are described. The boundaries of band transport in the insulating regime and metallic behavior are clarified with respect to AOS FETs.

## II. Evidence for band transport

Several studies have reported evidence of difficulty in reconciling with a purely hopping transport picture, but more consistent with trap-influenced band transport, in amorphous oxide semiconductors. [19,22] The evidence in support of band transport can be described as consisting of several separate pieces of evidence that, must be considered together. These are summarized below: This includes structural evidence of the morphology of these nominally amorphous semiconductor films that show nanoscale crystalline domains, electronic structure calculations and supporting experimental data, and charge transport including scattering mechanisms dominant in band transport and the behavior of carrier velocity at high electric fields. These aspects are discussed in this section and some of them, notably charge transport and carrier velocity, are discussed in additional detail in separate sections that follow.

### (i) Morphology of nominally amorphous semiconductors

This section will address aspects of the morphology of amorphous semiconductors related to structural ordering that have been reported in various experiments. Detecting structural order in amorphous semiconductors is inherently challenging because any ordered regions are typically only a few nanometers in extent, which is near the resolution limit or sensitivity limit of many structural probes. Consequently, the reported evidence is often best interpreted as indicating some degree of local order rather than conventional long-range crystallinity. Studies show that in several of these semiconductors (such as *a*-Si, IGZO, ZTO, etc.), the deposited films consist of nanocrystalline domains of a few nm extent. This is summarized in **Table I** together with a listing of the experimental techniques used in these reports. This range of crystalline domain sizes is order of the typical mean free path means, providing justification in treating these films as being ordered over distances of a few nm for charge transport analysis. The greater tolerance for bond angle distortions reported for amorphous oxide semiconductors compared to amorphous silicon suggests that there will be fewer defects in AOS thin-films compared to *a*-Si. [31] This is also supportive of using an effective medium approximation in treating the entire semiconductor film as possessing a uniform set of physical properties.

Most earlier work on depositing AOS films have used radio frequency sputtering or solution-based deposition methods together with some annealing. [4,32,33,34] There is also evidence of improved ordering in the semiconductor film if growth methods such as atomic layer deposition or plasma enhanced atomic layer deposition (PEALD) are employed in the deposition of very thin layers of AOS (a few nm thickness). [5,35] This will make the case for the use of band transport picture even stronger than for sputtered AOS films.

**Table I. Measured crystalline domain sizes in amorphous semiconductors including the experimental method used to measure the domain size.**

| Semiconductor | Measured crystalline Domain or range of order | Experimental Method | Ref |
|---|---|---|---|
| *a*-Si | 1.5 nm | FEM + TEM | [36] |
| *a*-Si | 3 nm | High-resolution TEM | [37] |
| *a*-ZTO | 1.5 nm ~ 3 nm | FEM | [38] |
| *a*-ZITO | 1.1 nm ~ 1.5 nm | STFEM | [39] |
| IGZO | nanometer size | NBED | [40] |
| IGZO | 2 ~ 5 nm | STEM | [41] |

*FEM = Fluctuation Electron Microscopy
*TEM = Transmission Electron Microscopy
*STFEM = Scanning Transmission Fluctuation Electron Microscopy
*NBED = Nanobeam electron diffraction
*STEM = Scanning Transmission Electron Microscopy

ii) Observation of normal Hall Effect:

Hall measurements on AOS have resulted in observations of a normal Hall effect, in which the sign of the Hall voltage is consistent with the carrier type (electron or hole) present. This behavior is similar to that seen in crystalline silicon and many other conventional crystalline semiconductors. [27,42,43] While the normal Hall effect is also observed in some semiconductors with hopping transport, it is usually associated with systems with a well-defined velocity vector (or equivalent, *k*-vector), which is consistent with band transport. **Figure 1**a shows the measured Hall voltage of solution processed Zinc Tin Oxide (ZTO) device as a function of DC magnetic field at 300 K. The Hall voltage showed a linear dependence on magnetic field up to ±9T, with the expected sign for electron transport (i.e., normal Hall effect) and indicating a sheet carrier concentration of $1.56 \times 10^{13}$ cm$^{-2}$ assuming that the Hall factor $r_H = 1$. Additional details of the ZTO device fabrication method are provided in Ref. [44]

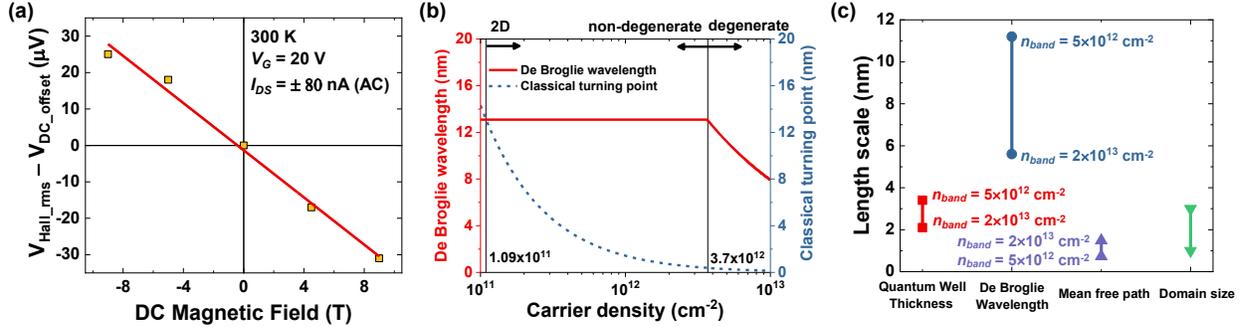

Figure 1. (a) Measured Hall voltage as a function of DC magnetic field for a ZTO TFT at room temperature. Carrier density is determined by $n = r_H I dB/(q dV)$. (b) De Broglie wavelength and Classical turning point distance as functions of total induced sheet carrier density, showing that for most carrier densities, the classical turning point is less than the de Broglie wavelength. (c) Illustration of the length scales of interest, including the ranges of effective quantum well thickness, De Broglie wavelength, and mean free path as a function of carrier density. The range of reported crystalline domain size of various oxide semiconductors is also shown for comparison. For calculations in (b) and (c), a nominal band mobility $\mu_0 = 20 \text{ cm}^2 V^{-1} s^{-1}$ and $m^*/m_0 = 0.34$ is assumed.

iii) Band structures have been reported by several groups and effective masses calculated theoretically as well as measured experimentally by angle resolved photoemission.[28,45,46] The effective masses and other key physical parameters for alloy AOS and associated binary semiconductors, from both calculations and measurements are summarized in **Table II**. The effective masses, which are below the free electron mass are consistent with reasonably dispersive conduction-band states and imply the existence of sufficiently wide bands and the likely prevalence of adequately high mobilities (> 10 cm$^2$/V-s). Such has been reported by many groups in experimental device structures. The calculated conduction band widths in AOS are much greater than both the thermal energy k$_B$T and phonon energies (which are of order 10's of meV), implying that broadening effects of conduction band energies due to these effects will be relatively small, which are also conditions for band transport to exist.[47]

iv) Experimentally measured mobility values for several AOS in FET structures are in the 10's of cm$^2$V$^{-1}$s$^{-1}$.[5,6,35,48,49,50] These mobilities correspond to mean fee paths > 1 nm. This is more than the value of typical lattice constants in single crystals of oxide semiconductors and is indicative of delocalized carriers and extended state transport. Mean free paths and their dependence on carrier density in prototypical AOS are described in a later section.

v) Carrier velocity values > $10^6$ cm/s have been reported for AOS based on FET studies. [25,51] Carrier velocities in systems with hopping transport, which typically exhibit a strong electric field dependence, are generally well below $10^6$ cm/s. [52] Carrier velocities of electrons that move in band states are ~ $4\times10^6$ cm/s at electric fields of $2\times10^5$ V/cm. Carrier velocities in amorphous IGZO FETs are described in additional detail in a later section.

Other favorable conditions for band transport that are present in AOS include a low density of traps compared to amorphous silicon. In conjunction with the effective mass values that are less than the free electron mass, $m_0$, the low trap density is conducive for the Fermi level to cross the band edge and also the mobility edge for carrier densities achievable in FET devices. All the above reasons make it not only justifiable but also appropriate to use a band transport rather than hopping to describe charge transport in AOS FETs.

**Table II. Key parameters of amorphous oxide semiconductors**

| Semiconductor | $m^*/m_0$ | $\varepsilon_s$ | $\varepsilon_\infty$ | $\hbar\omega_0$ (meV) |
|---|---|---|---|---|
| IGZO | 0.34 [45] | 8~11.5 [53,54] | 4 [53] | - |
| In$_2$O$_3$ | 0.14 ~ 0.55 [28] | 8.9 ~ 10.7 [28] | 3.82 ~ 4.1 [28] | 20 ~ 30 [29] |
| ZnO | 0.23 ~ 0.34 [28] | 7.5 ~ 8.9 [28] | 3.6 ~ 3.8 [28] | 72 [29] |
| SnO$_2$ | 0.12 ~ 0.3 [28] | 7, 9, 14 [28,30] | 3.7, 3.9 [28,30] | 33, 42, 83, 92 [30] |
| β-Ga$_2$O$_3$ | 0.2 ~ 0.4 [28] | 10 ~ 12 [28] | 2.9 ~ 4 [28] | 35 ~ 48 [30] |

\* $\varepsilon_s$ = static dielectric constants
\* $\varepsilon_\infty$ = High frequency dielectric constants
\* $\hbar\omega_0$ = longitudinal-optical phonon energy

### III. Evidence for Quasi 2D nature of accumulation channels in amorphous oxide semiconductor FETs

In a FET, field-induced charges are typically confined electrostatically at the interface between the gate insulator and semiconductor. In AOS FETs, an accumulation layer is usually formed. The dimensionality of such a system (whether quasi 2D or 3D) with electrostatically confined charges is a very important consideration for determining the density of states, other physical properties, and evaluating charge transport. The question of what makes a semiconductor inversion layer system 2-dimensional was examined by J. Robert Schrieffer [55], and is also summarized in the classic paper by Ando, Fowler, and Stern. [56] Schrieffer postulated in 1957 that a system can be considered as two-dimensional by comparing the De Broglie wavelength and the classical turning point distance. If the classical point distance is less or comparable to the De Broglie wavelength of an electron in the system, then a system can be considered 2-dimensional. This is because in such systems, electron motion perpendicular to the interface between the insulator and semiconductor (where charges are confined), cannot be treated classically.

The de Broglie wavelength is given by (for a 2D system):

$$\lambda = \frac{h}{\sqrt{2m^*E}} \quad (1)$$

where $E$ is the carrier energy, which in 2-dimensions is equal to $k_BT$ for non-degenerate systems or the Fermi energy ($E_F$) for degenerate systems. $m^*$ is the carrier effective mass. The classical turning point is given by $k_BT/qF_s$, where $F_s$ is the surface normal to the interface. $q$ is the element of charge, $T$ is the temperature, and $k_B$ is Boltzmann's constant. $F_s$ can be related to the induced charge density ($n_s$) by the relation:

$$F_s = \frac{qn_s}{\varepsilon_0\varepsilon_r} \quad (2)$$

For IGZO, the Schrieffer criterion is met for carrier densities $n_s > 1.11\times10^{11}$ cm$^{-2}$ as shown in **Fig. 1b**, which shows the De Broglie wavelength and classical turning point as function of carrier densities. For *a*-IGZO, the system can be regarded as quasi-2-dimensional when carrier density exceeds $1.1\times10^{11}$ cm$^{-2}$ assuming $m^*/m_0 = 0.34$ [45] and $\varepsilon_r = 10$ at 300 K. In addition, the system is non-degenerate for carrier densities below $3.7\times10^{12}$ cm$^{-2}$ and degenerate above this value. Such carrier densities are very easily achieved in AOS FETs, where typical carrier densities in the on-state of a FET are in the range $10^{12} - 10^{13}$ cm$^{-2}$ and higher. The estimated confinement length, which is much smaller than the de Broglie wavelength, and the observed single subband occupancy at carrier densities of interest in IGZO and ZTO are consistent with a quasi-2D description of the accumulation channel. Details are provided in the **Supplementary Materials S1**. These length scale ranges are summarized in **Fig. 1c**.

Apart from the classical turning point distance, it is useful to examine some other physical quantities that can corroborate the quasi-2D picture. These include the extent of energy level broadening due to scattering, the separation between energy levels, and whether only one sub-band is occupied. It may be noted that a band mobility of 100 cm$^2$V$^{-1}$s$^{-1}$, corresponds to a scattering time of $\sim1.9\times10^{-14}$ s for IGZO in the relaxation time approximation. The energy uncertainty associated with such scattering times is $\sim34$ meV, which is comparable to the thermal energy at room temperature. The Fermi energy corresponding to 34 meV for IGZO is $\sim 4.8\times10^{12}$ cm$^{-2}$, which is in the range of carrier densities in the on-state of an IGZO FET (which can often exceed $10^{13}$ cm$^{-2}$). These energy scales as well as the fact that for IGZO and ZTO only one subband is occupied under typical operating conditions (**Supplementary Materials S1**), indicate that it is justifiable to invoke a quasi-2D picture for charge transport, band structure, etc. Additional support for a 2D picture in AOS is provided by the observation of metal-insulator transitions at sheet conductivities characteristic of 2D systems and will be discussed in more detail in the following section.

### IV. Density of electronic states

In many earlier papers on AOS semiconductors, a 3D density of states was assumed. For reasons discussed above, it may be preferable to use a quasi 2D density of states. In the following, the construction of a density of states as a function of energy is described that considers both trap states and extended states separated by the mobility edge. The trap density of states is assumed to be exponential near the band edge, consistent with recent measurements.[17] The trap DOS that is deeper in energy has features that affect threshold voltage and residual carrier density.[8]

Disordered $a$-IGZO contains traps with a sample-dependent energy and density distribution below the band edge. Mobile charge carriers move in extended states in $a$-IGZO and are intermittently trapped. When in traps, carriers can move by hopping or multiple trap and release (MTR). In the MTR picture, trapped carriers can be thermally activated above the band edge. Several earlier works described hopping transport in IGZO using a variable range hopping model. [17,45] The prevalence of MTR compared to hopping is more likely in higher mobility semiconductors with room temperature mobilities > 10 cm$^2$V$^{-1}$s$^{-1}$.

In a simplified picture, the density of states (DOS) of shallow traps in $a$-IGZO can be described by an exponential tail extending from band edge into band the gap, and can be expressed as:

$$D_{trap}(E_k) = \frac{N_T}{k_B T_{ta}} e^{(E_k - E_c)/k_B T_{ta}} \ (E_k < E_c) \tag{3}$$

where $N_T$ is the total trap density, $T_{ta}$ is the characteristic temperature of the trap DOS, $E_k$ is the carrier energy, $E_c$ is the conduction band edge, and $k_B$ is the Boltzmann constant.

The DOS of extended states, in 2D or quasi-2D, is given by:

$$D_{band}(E_k) = \frac{gm^*}{2\pi\hbar^2} \ (E_k > E_c) \tag{4}$$

where $g$ is the spin degeneracy factor, $m^*$ is the effective mass of $a$-IGZO and $\hbar$ is the reduced Planck's constant. For IGZO, the value of $m^*$ is taken as 0.34$m_0$, where $m_0$ is the free electron mass. [45] The valley degeneracy factor is assumed to be 1. $g = 2$ in most cases for IGZO TFTs.

The values of $N_t$ and $T_{ta}$ can be estimated by fitting experimental FET output characteristics with a theoretical model. The values of $N_t$ and $T_{ta}$ will alter the balance between trapped charge density and free charge density in accordance with the above equations. In **Fig. 2a**, the DOS of trap states was fitted with an $N_t = 6.5 \times 10^{12}$ cm$^{-2}$ and $T_{ta}$ of 1150K. **Figure 2a** shows the density of states distribution combining both band states ($D_{band}$) and trap states ($D_{trap}$). At relatively low carrier density ($n \sim 1 \times 10^{12}$cm$^{-2}$), the Fermi level lies slightly above conduction band edge (7 meV). At higher carrier density (1×10$^{13}$ cm$^{-2}$), the Fermi level shifts well above the conduction band edge (70 meV).

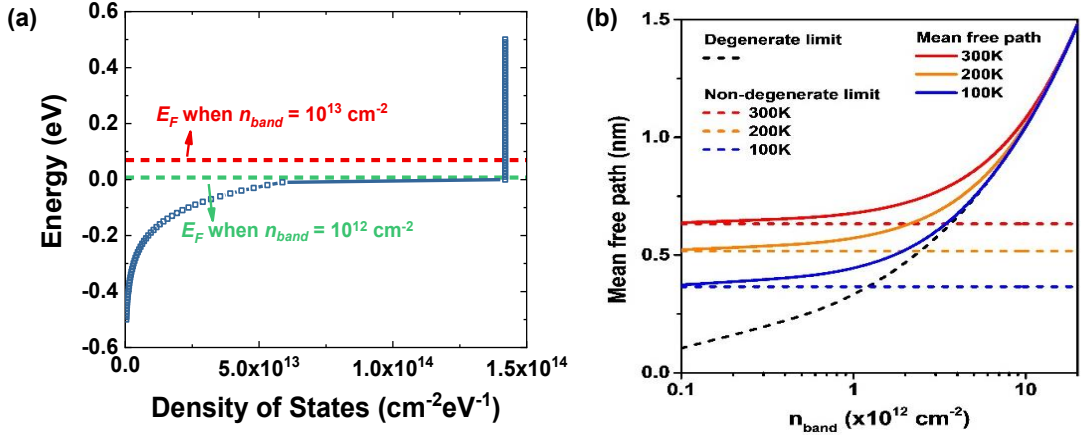

Figure 2. (a) Calculated density of states (DOS) as a function of energy. The band-state DOS is calculated as $1.42\times10^{14}$ cm$^{-2}$eV$^{-1}$ for $E > 0$ eV, while for $E < 0$ eV, the trap DOS is $6.5\times10^{13}\times e^{E/0.0991eV}$ cm$^{-2}$eV$^{-1}$ for $E < 0$ eV. The dashed lines indicate Fermi levels, $E_F$, corresponding to carrier densities of $n_{2D} = 10^{12}$ and $10^{13}$ cm$^{-2}$. (b) Calculated mean free path as a function of carrier density in the band at different temperatures. The colored horizontal dashed lines are the mean free path values in the non-degenerate limit based on Eq. 9. The colored solid lines are for the mean free path in more general conditions based on Eq. 10 and 11. A value $\mu_0 = 20$ cm$^2$/Vs for the nominal band mobility is assumed.

## V. Fermi level, Carrier density and Mean Free Path

In steady state, a part of the charge carriers induced in the channel by the gate occupy trap states while the rest are in extended states, where band transport can take place. The equilibrium populations of trapped and free carriers are conveniently described by a factor, designated as the MTR factor ($P_{MTR}$), obtained by integrating the DOS and the Fermi-Dirac distribution over energy: The MTR factor is the ratio of carrier density in the band to the total induced carrier density. [9]

$$n_{trap} = \int_{-\infty}^{E_c} \frac{N_t}{k_B T_{ta}} e^{(E_k - E_c)/k_B T_{ta}} f_{FD}\, dE_k \quad (5)$$

$$n_{band} = \int_{E_c}^{\infty} \frac{gm^*}{2\pi \hbar^2} f_{FD}\, dE_k \quad (6)$$

$$P_{MTR} = \frac{n_{band}}{n_{band} + n_{trap}} = \frac{n_{band}}{n_{total}} \quad (7)$$

where $f_{FD} = 1/\{1 + \exp[(E_k - E_f)/k_B T]\}$ is the Fermi-Dirac function, $T$ is the temperature, $E_f$ is the Fermi level, $n_{trap}$ is the trapped carrier density, $n_{band}$ is the density of carriers above the band, $n_{total}$ is the induced total carrier density.

In two dimensions, the mean free path can be estimated by equations that are described and listed below. In degenerate systems for electrons, in which the Fermi energy is greater than the average thermal energy of carriers (which is $\sim 2k_B T$, where $k_B$ is Boltzmann's constant and $T$ is the temperature), the mean free path $l$ is given by:

$$l = \frac{\hbar \mu_0}{e}\sqrt{2\pi n_{band}} \tag{8}$$

where $e$ is the element of charge, $\hbar$ is the reduced Planck's constant, $n_{band}$ is the sheet carrier density above band edge, and $\mu_0$ is the nominal band mobility. Parabolic bands are assumed, and the effective mass approximation is used. In non-degenerate electron gases, in which the Fermi energy is less than the average thermal energy, the mean free path can be estimated from:

$$l = \frac{\mu_0}{e}\sqrt{2k_B T m^*} \tag{9}$$

where $m^*$ is the carrier effective mass. In practice, however, the Fermi level in many cases lies close to the border between degenerate and non-degenerate conditions. In such cases, the mean free path is given by:

$$l = \frac{\mu}{e}\sqrt{2E_{av} m^*} \tag{10}$$

$$E_{av} = \frac{\int_{E_c}^{\infty} g(E)(-\partial f_{FD}/\partial E) E dE}{\int_{E_c}^{\infty} g(E)(-\partial f_{FD}/\partial E) dE} \tag{11}$$

where $E_c$ is the conduction band edge, $E$ is the carrier energy, $g(E)$ is the density of states (DOS), $f_{FD}$ is the Fermi-Dirac distribution functions. $E_{av}$ is the average energy of charge carriers for transport. Equations (8)-(10) can also be rewritten in terms of conductivity and carrier density instead of mobility. In **Fig. 2b**, the mean free path calculated from the equations listed above are plotted at different temperatures to help relate the mean free path to quantities such as carrier density and temperature. The nominal band mobility is set to be 20 cm$^2$/Vs and carrier effective mass is assumed to be 0.34$m_0$ ($m_0$ is the bare electron mass). Equations 8-10 are strictly valid in semiconductors with a well-defined energy-momentum dispersion, which is the case for semiconductors with carriers that are delocalized, in which the uncertainties in wave vector value are sufficiently small. In semiconductors with carriers delocalized over small distances, uncertainties in wave vector will translate to uncertainties in other quantities such as the mean free path.

## VI. Charge transport in AOS FETs

Early work on charge transport in amorphous oxide semiconductors generally considered hopping transport as the main charge transport mechanism.[17,18] With mobilities exceeding several 10's of cm$^2$V$^{-1}$s$^{-1}$ in many AOS including IGZO and ZTO[48,57], it is increasingly difficult to invoke hopping as the main mechanism. In earlier work, Wang and Dodabalapur proposed an extended multiple trap and release model as the dominant charge transport mechanism with hopping playing a lesser role, especially in high-mobility semiconductors.[58]

In MTR transport, charges exist in both trap states and extended states, with a distribution governed by a Fermi-Dirac distribution function. MTR is consistent with the density-of-states picture described above. MTR can explain several experimentally observed phenomena including temperature and carrier density dependence of mobility, observation of carrier velocities > $10^6$ cm/s, and evidence of soft velocity saturation at high electric fields (> $10^5$ V/cm). MTR transport can also been used for describing charge transport organic semiconductor FETs including nanoscale transistors. [59-61]

Most treatments of MTR transport assume that the nominal band mobility is a value $\mu_0$, and do not explicitly calculate this from fundamental considerations. It is necessary for a complete picture of transport in high mobility AOS to calculate the band mobility of electrons and to assess the contributions of various scattering mechanisms in determining the overall mobility. In previous work, our group has analyzed multiple scattering mechanisms in AOS such as ZTO and IGZO. These include trapped carrier scattering [58,62], phonon scattering mechanisms including surface phonons and polar optical phonons [62,63], and interface roughness scattering. In high mobility AOS transistors, the dominant mechanisms are trapped carrier scattering and optical phonon scattering. The overall nominal band mobility ($\mu_0$) is got from Matthiessen's rule and the mobilities due to the individual scattering mechanisms listed above. Trapped carrier scattering is to be expected in a disordered semiconductor with a significant density of trap states. The trapped carriers (which are localized charges) will scatter the charges that move in extended states. This Coulomb scattering mechanism is efficient since the trapped and band carriers are spatially co-located and the density of trapped carriers, with is of order $N_T$, can be quite large.

Exemplary mobility calculations are shown in **Figs. 3** for IGZO for the main scattering mechanisms, trapped carrier scattering and optical phonon scattering and the average band mobility as a function of temperature and carrier density. Also shown is the expected thin-film mobility that factors in charges in traps and a percentage of carriers with low mean free paths that are localized even when they occupy band states. This has also been described in several previous publications. [19,25,64] The details of these calculations, including mobilities due to various scattering mechanisms, are in **Supplementary Materials S2**. **Table IIIa** shows the inputs to the mobility calculations.

The mobility due to trapped carrier scattering exhibits a dependence on temperature and carrier density characteristic of Coulomb scattering, as shown in **Fig. 3a**. At low carrier density, the mobility increases with temperature due to increasing carrier energy with temperature which makes Coulomb scattering less effective. At higher carrier densities, degenerate conditions result in a very weak temperature dependence. There is still a strong carrier density dependence of mobility due to increased screening of the Coulomb potential at higher carrier densities. Due to the strong polar nature of AOS, polar optical phonon scattering is the dominant phonon scattering mechanism, and is especially dominant at higher temperatures. As expected, the mobility falls with increasing temperature with a tendency to saturate above about 300 K. When high k-dielectrics are used in AOS FETs, surface optical phonon (SO phonon) scattering becomes increasingly important although LO phonon scattering remains the dominant mechanism. Additionally, when the gate insulator-semiconductor interface has significant roughness, interface roughness scattering is enhanced and can be very important in lower mobility devices with rough interfaces. [64] **Figure 3c** shows the average band mobility obtained from the dominant scattering mechanisms using Matthiessen's rule. The combination of all the scattering mechanisms will result in a range of path lengths, some of which would be less than lattice constant. A transport reduction factor was introduced to account for this distribution of path lengths and its effect on mobilities.[19] It has a value of less than 1 and results in a decrease in mobility because of there being some carriers with path lengths smaller than the lattice constant and therefore localized. The transport reduction factor is a consequence of strong scattering that takes place dynamically in the semiconductor, with some of the stronger scattering events leading to temporal and spatial localization of carriers. The thin-film mobility, which is what is measured experimentally in a FET, is shown in **Fig. 3d**. It is nominal band mobility, $\mu_0$, reduced by three factors as shown in Eq. 8 **in Supplemental Materials S2**. These factors include the MTR factor and TRF factor described above and defined in **Supplemental Materials S2**. The thin-film mobility is expected to be both temperature and carrier density dependent.

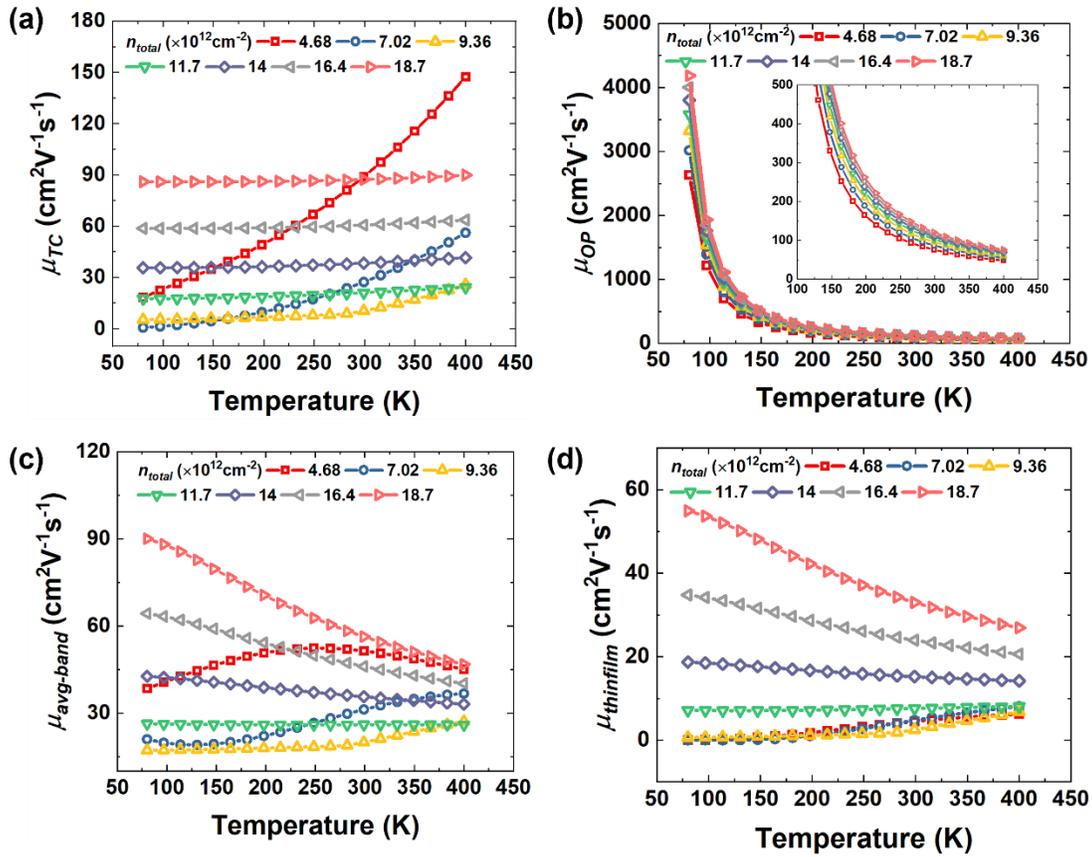

**Fig. 3. Temperature dependence of (a) $\mu_{TC}$, (b) $\mu_{OP}$, (c) $\mu_{avg-band}$, and (d) $\mu_{thin-film}$ for different sheet carrier densities. Calculations were conducted with $C_{ox}$ = 0.75 µF/cm², $V_{ov}$ = 1 V ~ 4 V, using $N_T$ = 6.5×10¹²/cm², $T_{ta}$ = 1150 K, $\varepsilon_s$ = 10, $\varepsilon_\infty$ = 4, and $\hbar\omega$ =30 meV**

The uncertainty in values of key parameters used in mobility calculations (**Table IIIa**) must be emphasized. More data is needed on effective mass and its variation with alloy composition as well as the low and high frequency dielectric constants. This difference is important in determining mobilities due to polar optical phonon scattering, expected to be the dominant phonon scattering mechanism in these polar semiconductors. The strong polar nature of $Ga_2O_3$ compared to GaN, for example, leads to lower electron mobilities in crystalline $Ga_2O_3$ due to stronger optical phonon scattering. In amorphous IGZO and related semiconductors, the phonon modes will be different from single crystals and more smeared out. The mobilities in single crystal IGZO FETs reported are only a factor of slightly (i.e., 1~2x) higher than those in amorphous IGZO. [65,66] The mobilities of electrons in crystalline IGZO, crystalline $In_2O_3$, and $Ga_2O_3$ are much lower than those of III-V compound semiconductors, which are less polar. This underlines the importance of polar optical phonon scattering in these oxide semiconductors at room temperature. In **ss**, longitudinal optical (LO) phonon energies are shown for IGZO and also crystalline binary oxide semiconductors.

VII. Spatial non-uniformity and percolation conduction

The likely spatial non-uniformity in alloy composition present in an AOS semiconductor film will result in spatially varying bandgaps. The scale of local spatial uniformity will depend on deposition methods and will set an upper limit on mean free path since scattering will result in the potential fluctuations present at domain boundaries. Since mean free paths are a few nm at best in these materials, compositional variations on a larger spatial scale than a few nm will result in the emergence of percolation pathways over regions of lower bandgap, as schematically illustrated in **Fig. 4a.** Percolation was proposed by Hosono and co-workers and has been invoked by various other groups. [67,68] The corresponding energy landscape associated with these spatial bandgap fluctuations is illustrated in **Fig. 4b.**

One favorable consequence of the formation of percolation pathways due to spatial bandgap variations is the charge focusing that occurs when the band carrier density is increased in the pathway regions, leading to higher mobilities and velocities, as can be discerned from **Fig. 3c** and **Fig. 3d**. The higher local mobilities are principally due to the enhanced screening of trapped carrier scattering at higher local carrier densities. While higher local carrier densities resulting from confinement of charge in these valleys may improve charge transport by raising the local Fermi energy relative to the band edge, any potential barriers will impede charge flow. Thus, the effects of percolation will be strongly sample dependent. It must be emphasized that percolation is a real-space effect and for a complete picture must be combined with the MTR transport picture described above. It is also likely that potential barriers that could interrupt percolation pathways will be less problematic in small geometry FETs with channel lengths < 50 nm, in which higher lateral electric fields are expected.

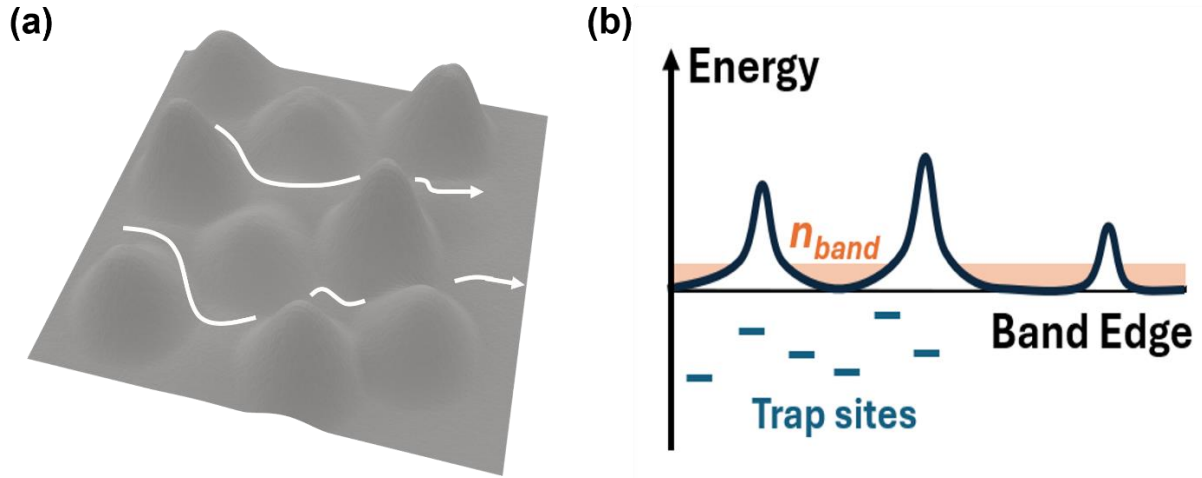

**Fig. 4 Schematic illustration of percolation conduction in disordered amorphous semiconductors: (a) real-space percolation pathways and (b) the corresponding energy landscape.**

The spatial scales over which percolation is expected to prevail will have the mean free path as the lower limit (few nm), The upper limit could be set by device channel length in BEOL FETs (< 100 nm). In recent work, large area IGZO FETs were probed by spatially resolved microwave impedance spectroscopy. [69,70] In the sub-threshold conduction regime, there is evidence of spatial non-uniformity of electrical conductivity with a characteristic length scale of about 200 nm that may be related to disorder, trapping, or possibly compositional variations. This provides one example of the upper limit of the length scale of non-uniformity in material composition or trap density, which is highly dependent on sample growth conditions.

Therefore, in a microscopic treatment of charge transport, the transport reduction factor (TRF), introduced above and in Refs [19,64] has to be incorporated into percolation theory, and will have a spatially dependent value. Spatial non-uniformities in composition and trap density will be reflected in a corresponding spatial dependence of band transport parameters. In potential valleys, scattering rates would be different compared to other regions, also contributing to the spatial variation in TRF. The TRF implicitly factors in non-uniformity in composition and/or trap density. These aspects need to be investigated in detail and a more detailed theoretical treatment of a spatially dependent TRF is in progress and will be reported elsewhere. A more detailed theoretical model will need to take such spatial non-uniformities into account.

### VIII. Carrier velocity

The carrier velocity at high electric fields and the velocity-electric field behavior provides additional insights into charge transport. Recent work by Yoon et al., on 50-100 nm channel length IGZO FETs that modeled experimental data taking into account thermal effects, contact resistance, position dependent electric fields in the channel, and a density of states and band transport picture described above [51]. A representative simulated $I_D$-$V_D$ data is shown in **Fig. 5a** and corresponding input parameters and extracted electrical properties are summarized in **Tables III**. For amorphous oxide semiconductor FETs, the average carrier velocity which is based on a modified Caughey-Thomas model can be described using equation: [25,71]

$$v = \frac{P_{MTR}P_{TRF}\mu_{avg-band}E}{[1+(\mu_{avg-band}E/v_{sat})^\beta]^{\frac{1}{\beta}}} \quad (12)$$

where $P_{MTR}$ represents the ratio of carriers in extended states to the total induced carriers, $P_{TRF}$ is a phenomenological transport reduction factor accounting for carriers with sufficiently long path length [19], $\mu_{avg-band}$ is average mobility of carriers in extended states, $E$ is electric field in the channel, $v_{sat}$ is saturation velocity of trap-free IGZO, and $\beta$ is a fitting parameter. The results have found that the average carrier velocity of all field-induced carriers (including both carriers in trap states and in band states) exceeds $2\times10^6$ cm/s at electric fields more than $2\times10^5$ V/cm. If current flow is attributed entirely to carriers in the band, then the velocity value exceeds $4\times10^6$ cm/s, as shown in **Fig. 5b**. The velocity is calculated from the charge transport and device model mentioned above and described in detail in **Supplementary Materials S3**. In addition to the input parameters needed for mobility calculations that are listed in **Table IIIa**, a few more parameters listed in **Table IIIb** need to be provided for calculating carrier velocity. The parameters that are obtained in the output from the model calculation include position dependent carrier mobility and carrier velocity as well as electric field. The velocity calculations presented in this paper assume there is no Joule heating. In earlier work Joule heating was included in the calculations as these can be significant in small geometry FETs at high gate and drain voltages. The change in carrier velocity due to Joule heating from 300 K to ~ 410 K is about 10%, being slightly smaller at higher temperatures due to phonon scattering.

**Table IIIa. Input parameters used for mobility calculations in a 50 nm channel length IGZO FET.**

| Input parameters for mobility calculations | | | |
|---|---|---|---|
| $N_T$ | $6.5\times10^{12}$ cm$^{-2}$ | $T_{ta}$ | 1150 K |
| $\varepsilon_s$ | 12 | $\varepsilon_\infty$ | 4 |
| $\hbar w$ | 20 meV | Lattice constant, $a$ | 0.33 nm |
| $t_{ox}$ | 9 nm | $\varepsilon_r$ | 7.5 |
| $V_{GS}$ | 4 V | $V_{DS}$ | 2.5 V |
| $V_T$ | -0.2 V | | |

**Table IIIb. Additional input parameters used for velocity calculations in a 50 nm channel length IGZO FET.**

| Additional input parameters for velocity and electric field calculations | | | |
|---|---|---|---|
| $L_{ch}$ | 50 nm | $W_{ch}$ | 3.4 µm |
| $R_c$ $(V_{GS}, V_{DS})^*$ | 1.94 kΩµm | | |

*$R_c$ at $V_{GS}$ = 4V and $V_{DS}$ = 2.5 V

**Table IIIc. Extracted electrical properties of 50 nm channel length IGZO FET obtained from model output.**

| | | | |
|---|---|---|---|
| $V_{ch(source)}$ | 0.92 V | $V_{ch(drain)}$ | 1.57 V |
| $E(y)$ (=$dV_{ch}/dL_{ch}$) | $2.38 \times 10^5$ V/cm | $\mu_{OP}(y)$ | 82 cm$^2$v$^{-1}$s$^{-1}$ |
| $\mu_{TC}(y)$ | 42.7 cm$^2$v$^{-1}$s$^{-1}$ | $\mu_0(y)$ | 28.1 cm$^2$v$^{-1}$s$^{-1}$ |
| $\mu_{avg-band}(y)$ | 36.1 cm$^2$v$^{-1}$s$^{-1}$ | $\mu_{thinfilm}(y)$ | 13.58 cm$^2$v$^{-1}$s$^{-1}$ |

*Specific values listed that are dependent on $y$ are for channel position adjacent to the drain

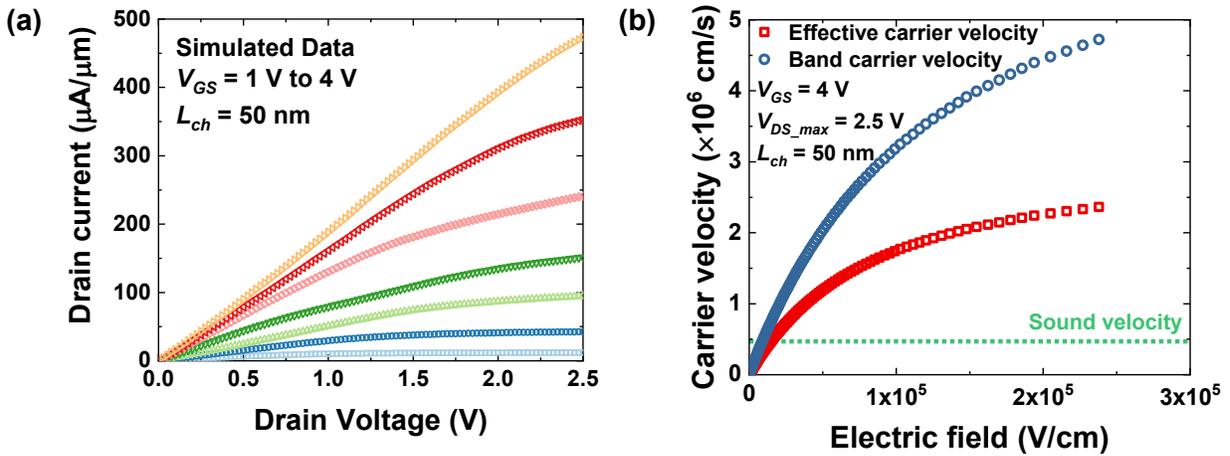

**Fig. 5** (a) Simulated $I_D$-$V_D$ characteristics of an IGZO FET with $L_{ch}$ = 50 nm at $V_{GS}$ = 1 – 4 V. (b) Effective and band carrier velocity as a function of local electric field ($E = dV_{ch}/dL_{ch}$) in IGZO FETs with channel length of 50 nm. The velocity of sound in IGZO, $4.7 \times 10^5$ cm/s (ref [72]), is included for comparison.

IX. **Relationship between Band Transport and Metal Insulator Transition**

The mobility edge, proposed by Sir Neville Mott [73], separates the extended states from localized states at T=0 K. At this temperature, it is also the minimum Fermi energy required for metallic behavior. At non-zero temperatures, the situation is more complex, especially in semiconductors such as AOS. It becomes necessary to separate the threshold energy for extended state transport (which is essentially the band edge) from that required for metallic behavior. The metal-insulator transition is defined by a threshold conductivity (in 2D) of $e^2/h$, which is ~ 38.7μS.[74] These are illustrated in **Figs. 6.** For disordered systems, it is helpful to represent this information in a plot in which the x-axis is conductivity and the y-axis is mean free path. This is shown in **Fig. 6b**. A conductivity of $e^2/h$ represents the boundary between metallic and insulating regimes. A line representing a mean free path of the typical lattice constant is shown in this figure. For mean free paths above this, delocalized transport can occur, while for smaller mean free path values, hopping transport takes place. A regime where both extended state transport and metallic behavior exists is shown. In this regime, $l > a$ and $\sigma > e^2/h$.

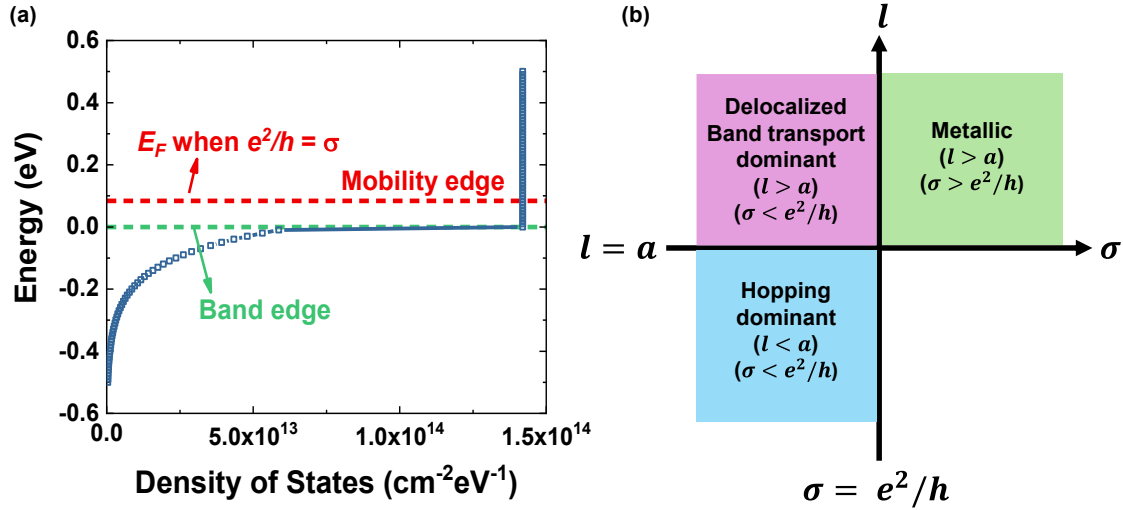

Fig. 6. (a) Calculated density of states (DOS) as a function of energy. The dashed lines indicate the band edge and the Fermi level at the metal-insulator transition criterion. The critical band carrier density is $1.2\times10^{13}$/cm$^2$, corresponding to $E_F = 84$meV. (b) Schematic classification of transport regime in terms of conductivity and mean free path, showing hopping, delocalized band transport, and metallic regimes.

The conductivity of AOS FETs at $T > 0$ K does show different temperature dependence at conductivities $> e^2/h$. This is seen in data from zinc tin oxide FETs as shown in **Figs. 7**. The thin-film mobility values appear to increase with decreasing temperature (below ~100 K) for high carrier densities. This behavior reflects the trend seen in the 2D conductivity above a conductivity of ~ 38 µS, which is $e^2/h$. Similar behavior has been reported for In$_2$O$_3$ FETs.[75]

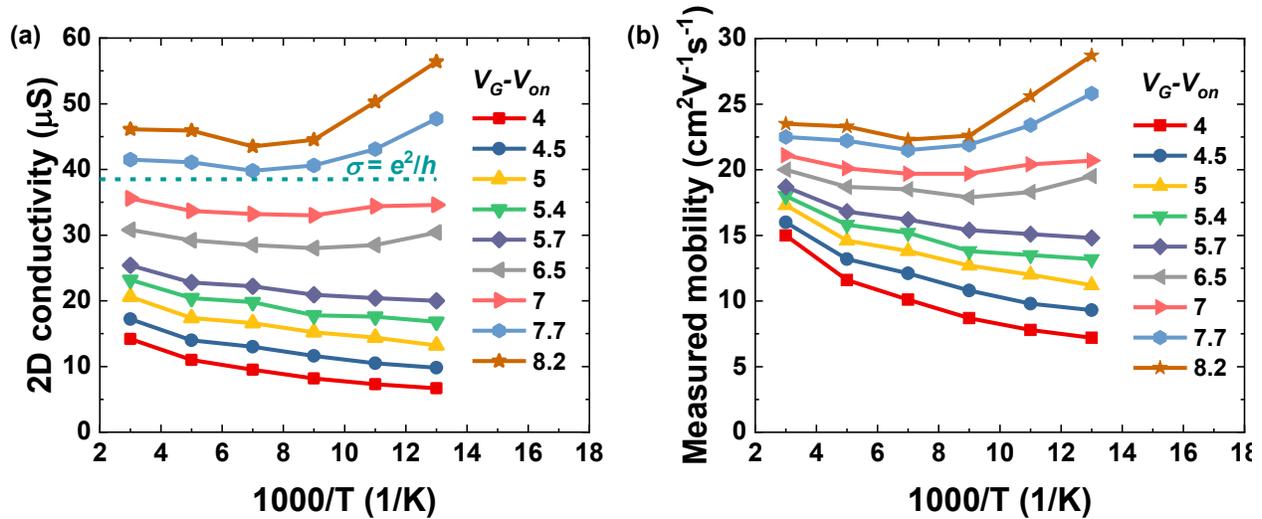

Fig. 7 (a) 2D conductivity of the ZTO TFT from Ref. [22]. The minimum 2D metallic conductivity is illustrated by the purple dashed line. (b) Measured mobility of this ZTO TFT as a function of gate voltage. $V_{on}$ is the on-voltage of the transistor [76] (analogous to the threshold voltage). For reference, the sheet carrier density, which increases linearly with $V_g - V_{on}$, is ~$1.15 \times 10^{13}$ cm$^{-2}$ for $(V_g - V_{on}) = 7.7$ V.

## Conclusions

The applicability to amorphous semiconductors of band-transport concepts originally developed for crystalline materials requires detailed justification, which has been the main focus of this paper. The combined experimental and theoretical evidence discussed here supports the use of such a picture for high-mobility amorphous oxide semiconductor (AOS) FETs. The main classes of evidence include structural observations, electronic-structure calculations and measurements, and charge-transport experiments together with quantitative transport analysis. Experimental evidence reported by several groups indicates nanometer-scale structural ordering in amorphous oxide semiconductors, suggesting that local order can persist over length scales of a few nanometers. This, together with the relative insensitivity of the conduction-band states to bond-angle distortions arising from their predominantly near-spherical *s*-orbital character, makes it reasonable to apply band-structure concepts over distances comparable to the transport mean free path.

Charge-transport evidence was interpreted within an extended multiple-trapping-and-release (MTR) framework. This framework is appropriate because it incorporates the effects of a trap density of states into a band-transport picture that also includes multiple scattering mechanisms. In high-mobility AOS FETs, the dominant scattering mechanisms are trapped-carrier scattering, a Coulomb scattering process analogous to charged-impurity scattering, and optical-phonon scattering. The measured mobilities, which are in the several tens of cm$^2$V$^{-1}$s$^{-1}$ range, are consistent with band transport. The carrier-density dependence of mobility, observed by multiple groups including our own, can be explained by screening within the MTR picture. The observation of carrier velocities exceeding $2 \times 10^6$ cm/s at high electric fields, together with the tendency toward velocity saturation, provides additional support for a band-transport picture. The extended MTR model can also incorporate real-space effects such as percolation conduction due to spatial inhomogeneities in material composition and/or trap density profile. The transport calculations were based on a quasi-2D density of states and corresponding device physics, justified by metrics such as the de Broglie wavelength, classical turning-point length, and mean free path. Finally, the relation between band transport and metallic conductivity in quasi-2D AOS FETs was discussed with reference to experimental data. Taken together, these results support the picture that charge transport in high-mobility AOS FETs is best described as trap-influenced transport through extended states in a quasi-2D channel.

**Acknowledgments**

The authors would like to thank Dr. Chenguan Lee, Prof. Leander Schulz, Prof. Keji Lai for helpful discussion. The work was supported by the National Science Foundation under Grant NNCI-2025227. Dr. Ananth Dodabalapur is grateful for support from Motorola Regents Chair Professorship in Electrical and Computer Engineering.

**References**

1. H.-M. Kim, J. Lee, S. Park, W. Choi, S. Kim, K.-S. Jeon, Y.-H. Choi, J.-S. Lee and J. Jang, SID Int. Symp. Dig. Tech. Pap. 51, 335–338 (2020)
2. Y.-M. Ha, S. K. Kim, H. Choi, S.-G. Lee, K.-S. Park and I. Kang SID Int. Symp. Dig. Tech. Pap. 47, 940-943 (2016)
3. Y. Takeda, S. Kobayashi, S. Murashige, K. Ito, I. Ishida, S. Nakajima, H. Matsukizono and N. Makita, SID Int. Symp. Dig. Tech. Pap. 50, 516-519 (2019)
4. C. Yoon, J. Ahn, Y. Zhou, J. P. Kulkarni and A. Dodabalapur, ACS nano **19** (43), 37712-37721 (2025).
5. J. Zhang, Z. Zhang, Z. Lin, K. Xu, H. Dou, B. Yang, X. Zhang, H. Wang and P. D. Ye In *First Demonstration of BEOL-Compatible Atomic-Layer-Deposited InGaZnO TFTs with 1.5 nm Channel Thickness and 60 nm Channel Length Achieving ON/OFF Ratio Exceeding 10 11, SS of 68 mV/dec, Normal-off Operation and High Positive Gate Bias Stability*, 2023 IEEE Symposium on VLSI Technology and Circuits (VLSI Technology and Circuits), IEEE: 2023; pp 1-2.
6. S. Datta, E. Sarkar, K. Aabrar, S. Deng, J. Shin, A. Raychowdhury, S. Yu and A. Khan, In *Amorphous oxide semiconductors for monolithic 3D integrated circuits*, 2024 IEEE Symposium on VLSI Technology and Circuits (VLSI Technology and Circuits), IEEE: 2024; pp 1-2.
7. A. Belmonte, H. Oh, S. Subhechha, N. Rassoul, H. Hody, H. Dekkers, R. Delhougne, L. Ricotti, K. Banerjee and A. Chasin, In *Tailoring IGZO-TFT architecture for capacitorless DRAM, demonstrating> 10 3 s retention,> 10 11 cycles endurance and L g scalability down to 14nm*, 2021 IEEE International Electron Devices Meeting (IEDM), IEEE: 2021; pp 10.6. 1-10.6. 4.
8. H.-H. Hsieh, T. Kamiya, K. Nomura, H. Hosono and C.-C. Wu, Applied Physics Letters **92** (13) (2008).
9. X. Wang and A. Dodabalapur, Journal of Applied Physics **132** (4) (2022).

10. S. Lee, A. Nathan, Y. Ye, Y. Guo and J. Robertson, Scientific reports **5** (1), 13467 (2015).
11. S. Lee and A. Nathan, Applied Physics Letters **101** (11) (2012).
12. Y. Luo and A. J. Flewitt, Physical Review B **109** (10), 104203 (2024).
13. H. U. Huzaibi, S.-T. Han and M. Zhang, npj Flexible Electronics (2025).
14. I. I. Fishchuk, A. Kadashchuk, A. Bhoolokam, A. de Jamblinne de Meux, G. Pourtois, M. Gavrilyuk, A. Köhler, H. Bässler, P. Heremans and J. Genoe, Physical Review B **93** (19), 195204 (2016).
15. Z. Zong, L. Li, J. Jang, N. Lu and M. Liu, Journal of Applied Physics **117** (21) (2015).
16. H. He, Y. Liu, B. Yan, X. Lin, X. Zheng and S. Zhang, IEEE Transactions on Electron Devices **64** (9), 3654-3660 (2017).
17. W. C. Germs, W. Adriaans, A. Tripathi, W. Roelofs, B. Cobb, R. Janssen, G. Gelinck and M. Kemerink, Physical Review B—Condensed Matter and Materials Physics **86** (15), 155319 (2012).
18. S. Lee, A. Nathan, J. Robertson, K. Ghaffarzadeh, M. Pepper, S. Jeon, C. Kim, I.-H. Song, U.-I. Chung and K. Kim, In *Temperature dependent electron transport in amorphous oxide semiconductor thin film transistors*, 2011 International Electron Devices Meeting, IEEE: 2011; pp 14.6. 1-14.6. 4.
19. X. Wang, L. F. Register and A. Dodabalapur, Physical Review Applied **11** (6), 064039 (2019).
20. Y. Luo and A. J. Flewitt, Journal of Non-Crystalline Solids **654**, 123436 (2025).
21. L. Li, N. Lu and M. Liu, IEEE Electron Device Letters **35** (2), 226-228 (2014).
22. C.-G. Lee, B. Cobb and A. Dodabalapur, Applied Physics Letters **97** (20) (2010).
23. X. Huang, C. Wu, H. Lu, F. Ren, D. Chen, R. Jiang, R. Zhang, Y. Zheng and Q. Xu, Solid-state electronics **86**, 41-44 (2013).
24. A. Bhoolokam, M. Nag, S. Steudel, J. Genoe, G. Gelinck, A. Kadashchuk, G. Groeseneken and P. Heremans, Japanese Journal of Applied Physics **55** (1), 014301 (2016).
25. X. Wang and A. Dodabalapur, IEEE Transactions on Electron Devices **68** (1), 125-131 (2020).
26. M. Cai and R. Yao, Solid-State Electronics **141**, 23-30 (2018).
27. W. Wang, G. Xu, M. D. H. Chowdhury, H. Wang, J. K. Um, Z. Ji, N. Gao, Z. Zong, C. Bi and C. Lu, Physical Review B **98** (24), 245308 (2018).
28. J. A. Spencer, A. L. Mock, A. G. Jacobs, M. Schubert, Y. Zhang and M. J. Tadjer, Applied Physics Reviews **9** (1) (2022).
29. Y. Hu, J. Hwang, Y. Lee, P. Conlin, D. G. Schlom, S. Datta and K. Cho, Journal of Applied Physics **126** (18) (2019).
30. N. Ma, N. Tanen, A. Verma, Z. Guo, T. Luo, H. G. Xing and D. Jena, Applied Physics Letters **109** (21) (2016).
31. K. Nomura, H. Ohta, A. Takagi, T. Kamiya, M. Hirano and H. Hosono, nature **432** (7016), 488-492 (2004).
32. C.-G. Lee, S. Dutta and A. Dodabalapur, IEEE electron device letters **31** (12), 1410-1412 (2010).
33. S. R. Thomas, P. Pattanasattayavong and T. D. Anthopoulos, Chemical Society Reviews **42** (16), 6910-6923 (2013).
34. C. Wang, A. Kumar, K. Han, C. Sun, H. Xu, J. Zhang, Y. Kang, Q. Kong, Z. Zheng, Y. Wang and X. Gong, In *Extremely scaled bottom gate a-IGZO transistors using a novel patterning technique achieving record high G m of 479.5 μs/μm (V DS of 1 V) and f T of 18.3 GHz (V DS of*

*3 V)*, 2022 IEEE Symposium on VLSI Technology and Circuits (VLSI Technology and Circuits), IEEE: 2022; pp 294-295.
35. J. Sheng, T. Hong, H.-M. Lee, K. Kim, M. Sasase, J. Kim, H. Hosono and J.-S. Park, ACS applied materials & interfaces **11** (43), 40300-40309 (2019).
36. P. M. Voyles and J. R. Abelson, Solar energy materials and solar cells **78** (1-4), 85-113 (2003).
37. J. Phillips, J. Bean, B. Wilson and A. Ourmazd, Nature **325** (6100), 121-125 (1987).
38. E. Rucavado, Q. Jeangros, D. F. Urban, J. Holovský, Z. Remes, M. Duchamp, F. Landucci, R. E. Dunin-Borkowski, W. Körner and C. Elsässer, Physical Review B **95** (24), 245204 (2017).
39. A. Yan, T. Sun, K. B. Borisenko, D. B. Buchholz, R. P. Chang, A. I. Kirkland and V. P. Dravid, Journal of Applied Physics **112** (5) (2012).
40. N. Sorida, M. Takahashi, K. Dairiki, S. Yamazaki and N. Kimizuka, Japanese Journal of Applied Physics **53** (11), 115501 (2014).
41. R. Khan, M. Ohtaki, S. Hata, K. Miyazaki and R. Hattori, Nanomaterials **11** (6), 1547 (2021).
42. J. Socratous, S. Watanabe, K. K. Banger, C. N. Warwick, R. Branquinho, P. Barquinha, R. Martins, E. Fortunato and H. Sirringhaus, Physical Review B **95** (4), 045208 (2017).
43. T. Kamiya and H. Hosono, NPG Asia Materials **2** (1), 15-22 (2010).
44. L. Schulz, E.-J. Yun and A. Dodabalapur, Applied Physics A **115** (4), 1103-1107 (2014).
45. A. Takagi, K. Nomura, H. Ohta, H. Yanagi, T. Kamiya, M. Hirano and H. Hosono, Thin solid films **486** (1-2), 38-41 (2005).
46. V. Scherer, C. Janowitz, A. Krapf, H. Dwelk, D. Braun and R. Manzke, Applied Physics Letters **100** (21) (2012).
47. S. H. Glarum, Journal of Physics and Chemistry of Solids **24** (12), 1577-1583 (1963).
48. C.-G. Lee and A. Dodabalapur, Applied Physics Letters **96** (24) (2010).
49. H. Chiang, J. Wager, R. Hoffman, J. Jeong and D. A. Keszler, Applied Physics Letters **86** (1) (2005).
50. E. Fortunato, L. Pereira, P. Barquinha, A. M. Botelho do Rego, G. Gonçalves, A. Vilà, J. R. Morante and R. F. Martins, Applied Physics Letters **92** (22) (2008).
51. C. Yoon, X. Wang, J. V. Singh, S. K. Banerjee and A. Dodabalapur, arXiv preprint arXiv:2604.21225 (2026).
52. P. M. Borsenberger, W. T. Gruenbaum and L. J. S. Zumbulyadis, Japanese journal of applied physics **34** (12A), L1597 (1995).
53. S. Matsuda, T. Hiramatsu, R. Honda, D. Matsubayashi, H. Tomisu, Y. Kobayashi, K. Tochibayashi, R. Hodo, H. Fujiki and Y. Yamamoto, In *30-nm-channel-length c-axis aligned crystalline In-Ga-Zn-O transistors with low off-state leakage current and steep subthreshold characteristics*, 2015 Symposium on VLSI Technology (VLSI Technology), IEEE: 2015; pp T216-T217.
54. H.-J. Chung, J. H. Jeong, T. K. Ahn, H. J. Lee, M. Kim, K. Jun, J.-S. Park, J. K. Jeong, Y.-G. Mo and H. D. Kim, Electrochemical and Solid-State Letters **11** (3), H51-H54 (2008).
55. J. Schrieffer, Semiconductor surface physics, 55-69 (1957).
56. T. Ando, A. B. Fowler and F. Stern, Reviews of Modern Physics **54** (2), 437 (1982).
57. S. Samanta, K. Han, C. Sun, C. Wang, A. Kumar, A. V.-Y. Thean and X. Gong, IEEE Transactions on Electron Devices **68** (3), 1050-1056 (2021).
58. X. Wang and A. Dodabalapur, Annalen der Physik **530** (12), 1800341 (2018).
59. G. Horowitz, Advanced materials **10** (5), 365-377 (1998).

60. A. Dodabalapur, H. E. Katz and L. Torsi, (Google Patents, 1997).
61. L. Wang, D. Fine, T. Jung, D. Basu, H. von Seggern and A. Dodabalapur, Applied physics letters **85** (10), 1772-1774 (2004).
62. S. Lisesivdin, A. Yildiz, N. Balkan, M. Kasap, S. Ozcelik and E. Ozbay, Journal of Applied Physics **108** (1) (2010).
63. K. A. Stewart and J. F. Wager, Journal of the Society for Information Display **24** (6), 386-393 (2016).
64. X. Wang and A. Dodabalapur, Journal of Applied Physics **130** (14) (2021).
65. A. V. Glushkova, H. F. Dekkers, M. Nag, J. I. del Agua Borniquel, J. Ramalingam, J. Genoe, P. Heremans and C. Rolin, ACS Applied Electronic Materials **3** (3), 1268-1278 (2021).
66. S. Jeong, S. Jang, H. Han, H. Kim and C. Choi, Journal of Alloys and Compounds **888**, 161440 (2021).
67. T. Kamiya, K. Nomura and H. Hosono, Journal of display technology **5** (12), 462-467 (2009).
68. S. Lee, K. Ghaffarzadeh, A. Nathan, J. Robertson, S. Jeon, C. Kim, I.-H. Song and U.-I. Chung, Applied Physics Letters **98** (20) (2011).
69. J. Yu, Y. Zhou, X. Wang, A. Dodabalapur and K. Lai, Nano Letters **23** (24), 11749-11754 (2023).
70. J. Yu, Y. Zhou, X. Wang, X. Ma, A. Dodabalapur and K. Lai, ACS Applied Electronic Materials **6** (11), 8448-8454 (2024).
71. F. Assaderaghi, D. Sinitsky, J. Bokor, P. K. Ko, H. Gaw and C. Hu, IEEE Transactions on Electron Devices **44** (4), 664-671 (1997).
72. T. Yoshikawa, T. Yagi, N. Oka, J. Jia, Y. Yamashita, K. Hattori, Y. Seino, N. Taketoshi, T. Baba and Y. Shigesato, Applied Physics Express **6** (2), 021101 (2013).
73. N. Mott, Journal of Physics C: Solid State Physics **20** (21), 3075-3102 (1987).
74. S. Das Sarma and E. Hwang, Physical Review B **89** (23), 235423 (2014).
75. W. Xie, X. Zhang, C. Leighton and C. D. Frisbie, Advanced Electronic Materials **3** (3) (2017).
76. D. Hong, G. Yerubandi, H. Chiang, M. Spiegelberg and J. Wager, Critical Reviews in Solid State and Materials Sciences **33** (2), 101-132 (2008).

Supplementary Materials

# Physical Basis for Band Transport and Dimensionality in Amorphous Oxide Semiconductor Field-Effect Transistors

Ananth Dodabalapur,[1,2,*] Chankeun Yoon,[1,2] and Xiao Wang [1,2]

[1] *Chandra Family Department of Electrical and Computer Engineering, The University of Texas at Austin, Austin, Texas 78712, USA*

[2] *Microelectronics Research Center, The University of Texas at Austin, Austin, Texas 78758, USA*

\* *Author to whom correspondence should be addressed:* ananth.dodabalapur@engr.utexas.edu

## S1. Finite Barrier Schrödinger-Poisson Analysis Supporting Quasi-2D Approximation

To assess the validity of a quasi-2D description in IGZO, a finite-barrier and self-consistent Schrödinger-Poisson analysis was performed for IGZO/$Al_2O_3$ interface. The parameters used in the analysis are summarized in **Table S1**.

Table S1. Parameters used in the finite-barrier Schrödinger–Poisson analysis.

| Parameter | Value / Description |
|---|---|
| Semiconductor | IGZO accumulation layer |
| Gate dielectric | $Al_2O_3$ |
| Sheet carrier density ($n_s$) | $10^{13}$ cm$^{-2}$ |
| Temperature ($T$) | 300 K |
| IGZO effective mass ($m^*$) | 0.34 $m_0$ |
| IGZO dielectric constant | 10 |
| $Al_2O_3$ dielectric constant | 8 |
| Conduction band offset | 2.4 eV |
| Barrier | Finite barrier at IGZO/$Al_2O_3$ interface |
| Electrostatic | Self-consistent 1D Schrödinger-Poisson treatment |

Table S2. Calculated quantities from the finite-barrier Schrödinger–Poisson analysis.

| Output | Approximate value |
|---|---|
| Band bending at interface | 0.45 V |
| Lowest subband energy ($E_1$) | −0.181 eV |
| Second subband energy ($E_2$) | −0.0865 eV |
| Fermi level ($E_F$) | −0.122 eV |
| Subband spacing, ($E_2$–$E_1$) | 95 meV |
| Fermi level above $E_1$ | 59 meV |

\* Energies are referenced to bulk IGZO conduction band edge far from the interface.

Taking bulk IGZO conduction band edge far from the interface as the energy reference, the finite barrier and self-consistent Schrödinger-Poisson analysis yields an interface band bending of approximately 0.45 eV at $n_s = 10^{13}$cm$^{-2}$, lowest subband energy ($E_1$) ≈ −0.181 eV and second subband energy ($E_2$) ≈ −0.0865 eV. The Fermi level ($E_F$) is determined from the total subband occupancy by required to reproduce sheet carrier density ($n_s$), giving $E_F$ ≈ −0.122 eV. Since $E_F$ remains below $E_2$, the electron population is dominated by the first subband ($E_1$), while occupation of the second subband ($E_2$) is weak.

## S2. Various Mobility Calculations in AOS FETs

Carriers occupying extended states in AOS FETs are influenced by several scattering mechanisms. In this work, two dominant scattering mechanisms are considered: trapped carrier (TC) scattering and polar optical phonon (PO) scattering. Trapped carrier scattering originates from Coulomb interactions between mobile carriers and immobile charged carriers. The trapped carrier limited mobility ($\mu_{TC}$) is therefore calculated based on Coulomb scattering by immobile charged carriers and is given by [1,2]

$$\mu_{TC} = \frac{8\pi\hbar^3(\varepsilon_0\varepsilon_s)^2 k^3 d}{e^3 m^{*2} n_{trap}} \int_0^\pi \frac{\sin^2\theta}{(\sin\theta+\beta)^2} d\theta, \quad (1)$$

where $\beta = S_0/2k$, $S_0$ is the screening constant, $\varepsilon_s$ is static dielectric constant of amorphous oxide semiconductor, $k$ is the wave vector, $\theta$ is the scattering angle and $d$ is effective channel thickness obtained from quantum well approximation. Under two-dimensional conditions, the screening parameter and wave vector are calculated separately for non-degenerate and degenerate conditions:

$$S_0 = \frac{e^2 n_{band}}{2\varepsilon_0\varepsilon_s k_B T}, \quad k = \frac{\sqrt{2k_B T m^*}}{\hbar} \quad \text{(non-degenerate)}, \quad (2)$$

$$S_0 = \frac{2e^2 m^*}{4\pi\varepsilon_s \hbar^2}, \quad k = \sqrt{2\pi n_{band}} \quad \text{(degenerate)}. \quad (3)$$

The polar optical phonon limited mobility ($\mu_{OP}$) is written as [1,2]

$$\mu_{OP} = \frac{4\pi\hbar^2 \varepsilon_0 \varepsilon_\infty \varepsilon_s}{e\omega dm^{*2}(\varepsilon_s - \varepsilon_\infty)}[\exp(\frac{\hbar\omega}{k_B T}) - 1], \tag{4}$$

where $\varepsilon_\infty$ is high-frequency dielectric constant of amorphous oxide semiconductor and $\omega$ is polar optical phonon frequency. The mobility of carriers in extended states ($\mu_0$) is then obtained from Matthiessen's rule:

$$\frac{1}{\mu_0} = \frac{1}{\mu_{TC}} + \frac{1}{\mu_{OP}} \tag{5}$$

To describe charge transport within extended states, carriers with a path length ($l$) shorter than lattice constant ($a$) (i.e., $l < a$) are regarded as effectively localized, whereas those with $l > a$ are assumed to contribute to band transport. The path length of carriers in extended states is assumed to follow a statistical distribution with an average value equal to mean free path ($l_{mfp}$). [3] The mean free path is defined as [4]:

$$l_{mfp} = \begin{cases} \frac{\mu_0}{e}\sqrt{2k_B T m^*} & (\text{non} - \text{degenerate}), \\ \frac{\hbar\mu_0}{e}\sqrt{2\pi n_{band}} & (\text{degenerate}). \end{cases} \tag{6}$$

To account for the fact that not all carriers in extended states contribute equally to charge transport due to scattering, a phenomenological transport reduction factor, $P_{TRF}$, is introduced [3]. This factor is defined as the probability that a carrier travels a distance equal or greater than lattice constant ($a$), based on statistical distribution of path length [3,5]. Assuming an exponential distribution for free path lengths with a mean value of $l_{mfp}$, the $P_{TRF}$ is given by

$$P_{TRF} = \exp(-a/l_{mfp}). \tag{7}$$

By combining the multiple and trap release factor ($P_{MTR}$) (Eq. 7 in the main manuscript), with the transport reduction factor ($P_{TRF}$), the overall thin-film mobility of the AOS FET can be expressed as

$$\mu_{thin\,film} = P_{MTR}P_{TRF}\mu_{avg-band} = P_{MTR}P_{TRF}(1 - \ln P_{TRF})\mu_0, \tag{8}$$

where $\mu_{avg-band} = (1 - \ln P_{TRF})\mu_0$.

### S3. Carrier Velocity Calculations in AOS FETs

Carriers occupying extended states gain energy from the lateral electric field, and velocity is eventually limited by optical phonon emission, resulting in velocity saturation during band transport. In AOS FETs, the carrier velocity is determined not only by the transport velocity in extended states, but also by the fraction of carriers that occupy those states. The ensemble carrier velocity of AOS FETs can be expressed as:

$$v = \frac{P_{MTR} P_{TRF} \mu_{avg-band} E}{[1+(\mu_{avg-band} E/v_{sat})^\beta]^{\frac{1}{\beta}}},  \qquad (9)$$

where $v_{sat}$ is the saturation velocity of trap-free AOS, and $\beta$ is fitting parameter.

**References**

1. Xiao Wang and Ananth Dodabalapur, IEEE Transactions on Electron Devices **68** (1), 125 (2020).
2. SB Lisesivdin, A Yildiz, N Balkan, MEHMET Kasap, SÜLEYMAN Ozcelik, and E Ozbay, Journal of Applied Physics **108** (1) (2010).
3. Xiao Wang, Leonard F Register, and Ananth Dodabalapur, Physical Review Applied **11** (6), 064039 (2019).
4. Xiao Wang and Ananth Dodabalapur, Journal of Applied Physics **132** (4) (2022).
5. Robert N Varney, American Journal of Physics **39** (5), 534 (1971).